# One More Tool for Understanding Resonance

Emanuel Gluskin (EE. Dept., Holon Institute of Technology, email: gluskin@ee.bgu.ac.il),
Doron Shmilovitz (Eng. Faculty, Tel-Aviv University, email: shmilo@eng.tau.ac.il),
Yoash Levron (Eng. Faculty, Tel-Aviv University, email:  yoash_lev@yahoo.com)

**Abstract:**  We propose the application of graphical convolution to the analysis of the resonance phenomenon.  This time-domain approach encompasses both the finally attained periodic oscillations and the initial transient period.  It also provides interesting discussion concerning the analysis of non-sinusoidal waves, based not on frequency analysis, but on direct consideration of waveforms, and thus presenting an introduction to Fourier series.  Further developing the point of view of graphical convolution, we come to a new definition of resonance in terms of time domain.

## 1. Introduction

### 1.1 General

The following material fits well into an "Introduction to Linear Systems", or "Mechanics", and is relevant to a wide range of technical and physics courses, since the resonance phenomenon has long interested physicists, mathematicians, chemists, engineers, and, nowadays, also biologists.

The complete resonant response of an initially unexcited system has two different, distinguishable parts, and there are, respectively, two basic definitions of resonance, significantly distanced from each other.

In the widely adopted textbook [1] written for physicists, resonance is defined as a *linear increase of the amplitude of oscillations in a lossless oscillatory system*, obtained when the system is pumped with energy by a sinusoidal force at the correct frequency. Figure 1 schematically shows the "envelope" of the resonant oscillations being developed.

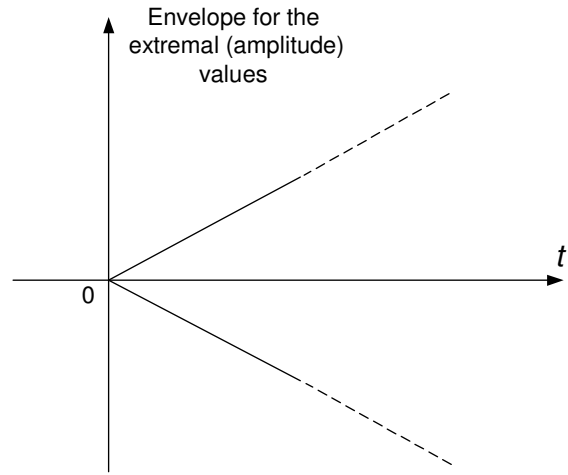

Fig. 1: The definition of resonance [1] as *linear increase of the amplitude*. (The oscillations fill the angle of the envelope.).  The *infinite* process of increase of the amplitude is obtained because of the assumption of *losslessness* of the system.

   Thus, a lossless system under resonant excitation absorbs more and more energy, and a steady state is never reached.  In other words, in the lossless system, the amplitude of the steady state and the 'quality factor' $Q$ (having a somewhat semantic meaning in such a system) are infinite at resonance.
   However, the slope of the envelope is always finite.  It depends on the amplitude of the input function, and not on $Q$.  Though the steady state response will never be reached in an ideal lossless system, the linear increase in amplitude by itself has an important sense.  When a realistic physical system absorbs energy resonantly, say in the form of photons of electromagnetic radiation, there indeed is some period (still we can ignore power losses, say, some back radiation) during which the system's energy increases linearly in time.  The energy absorption is immediate upon appearance of the influence, and the rate of the absorption directly measures the intensity of the input.
   One notes that the energy pumping into the system at the initial stage of the resonance process readily suggests that the sinusoidal waveform of the input function is not necessary for resonance; it is obvious (think, e.g., about swinging a swing by kicking it) that the energy pumping can occur for other input waveforms as well.  This is a heuristically important point of the definition of [1].
   The physical importance of the initial increase in oscillatory amplitude is associated not only with the energy pumping; the informational meaning is also important.  Assume, for instance, that we speak about the *start* of oscillations of the spatial positions of the atoms of a medium, caused by an incoming electromagnetic wave.  Since this start is associated with the *appearance* of the wave, it can be also associated with *the registration of a signal*.   Later on, the *established* steady-state oscillations (that are associated, because of the radiation of the atoms, with the refraction factor of the medium) influence the velocity of the electromagnetic wave in the medium.  As [2] stresses, -- even if this velocity is larger than the velocity of light (for refraction factor $n < 1$, i.e. when the frequency of the incoming wave is *slightly* higher than that of the atoms-oscillators), -- this does not contradict the theory of relativity, because there is already no signal.  Registration of any signal and its group velocity is associated with a (forced) transient process.

   A more pragmatic argument for the importance of analysis of the initial transients is that for any application of a steady state response, especially in modern electronics, we have to know how much time is needed for it to be attained, and this relates, in particular, to the resonant processes. This is relevant to the frequency range in which the device has to be operated.

   Contrary to [1], in textbooks on the theory of electrical circuits (e.g. [3-5]) and mechanical systems, resonance is defined as the *established* sinusoidal response with a relatively high amplitude proportional to $Q$. Only this definition, directly associated with *frequency domain analysis*, is widely accepted in the engineering sciences. According to this definition, the envelope of the resonant oscillations (Fig.2) looks even simpler than in Fig 1; it is given by two horizontal lines. This would be so for *any* steady-state oscillations, and the uniqueness is just by the fact that the oscillation amplitude is proportional to $Q$.

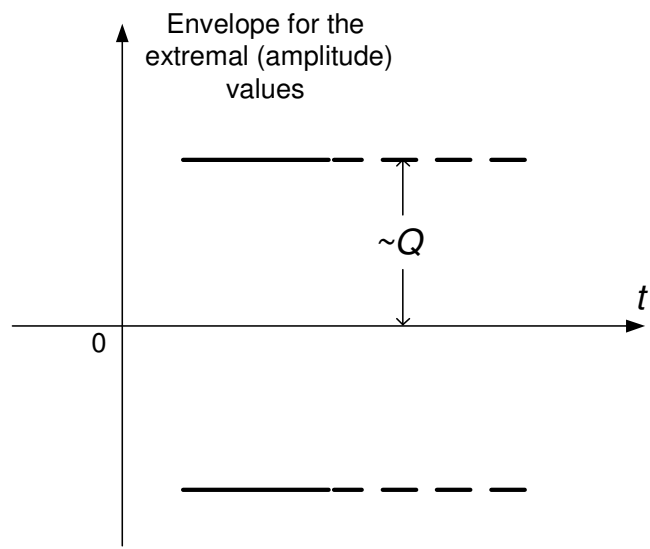

Fig. 2: The envelope of resonant oscillations, according to the definition of resonance in [3-4] and many other technical textbooks. When this steady state is attained?

   After being attained, the steady state oscillations continue "forever", and the parameters of the 'frequency response' can be thus relatively easily measured. Nevertheless, the simplicity of Fig. 2 is a seeming one, because it is not known *when* the steady amplitude becomes established, and, certainly, the 'frequency response' is *not* an immediate response to the input signal.

   Thus, we do not know via the definition of [1] when the slope will finish, and we do not know via the definition of [3-5] when the steady state is obtained.

   We shall call the definition of [1], "the '$Q$-$t$' definition", since the value of $Q$ can be revealed via *duration* of the initial/transient process in a real system. The commonly used definition [3-5] of resonance in terms of the parameters of the sustained response, will be called "the '$Q$-$a$' definition", where '$a$' is an abbreviation for "amplitude".

   Figure 3 illustrates the actual development of resonance in a second order circuit. The damping parameter $\gamma$ will be defined in Section 1.2.

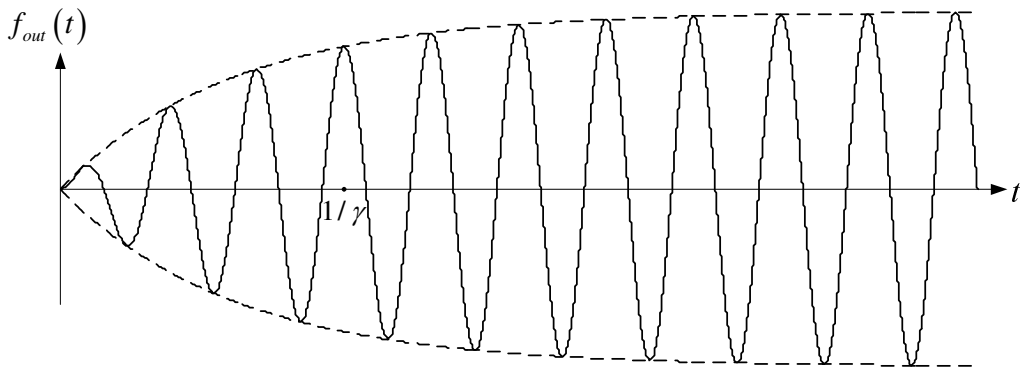

Fig.3: The illustration, for $Q = 10$, of the resonant response of a second-order circuit. Note that we show a case when the excitation is precisely at the resonant frequency, and the notion of "purely resonant oscillations" applies here to the *whole* process, and not only to the final steady stat*e* part.

The *Q-t* and *Q-a* parts of the resonant oscillations are well seen. For such a not very high $Q$ (i.e. $1/\gamma$ not much larger than the period of the oscillations) the period of fair initial linearity of the envelope includes only some half period of oscillations, but for a really high $Q$ it can include many periods. The *whole curve* shown is the resonant response. This response can be obtained when the external frequency is closing the self-frequency of the system, from the beats of the oscillations (analytically explained by the formulae found in Section 3) shown in Fig. 4.

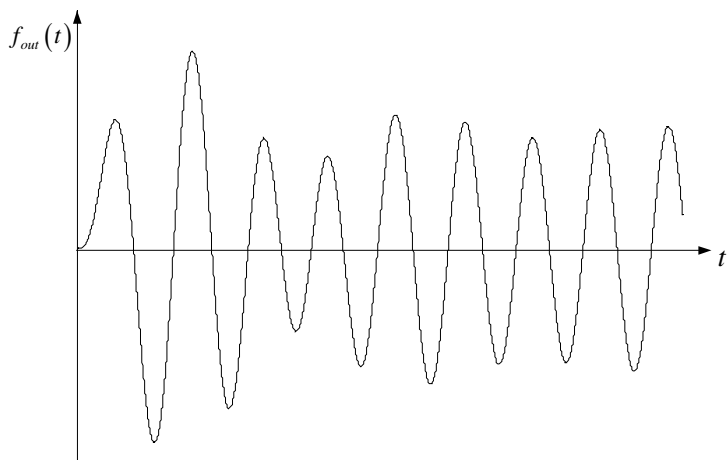

Fig. 4: Possible establishing of the situation shown in Fig. 3 through beats while adjustment the frequency. We can interpret resonance as "filtration" of the beats when the resonant frequency is found.

Note that the usual interpretation is somewhat different. It just says that the linear increase of the envelope, shown in Fig.1 can be obtained from the first beat of the *periodic* beats observed in a *lossless* system. Contrary to that, we observe the beats in a system with losses, and after adjustment of the external frequency obtain the *whole* resonant response shown in Fig. 3.

Our treatment of the topic of resonance for teaching purposes is composed of three main parts shown in Fig. 5. The first part briefly recalls traditional 'phasor' material relevant only to the *Q-a* part, which is necessary for introduction of the notations. The next part includes some simple *though usually omitted* arguments showing why the phasor analysis is insufficient. Finally, the third part includes the new tool which is complementary to the classical approach of [1], and leads to a nontrivial generalization of the concept of resonance.

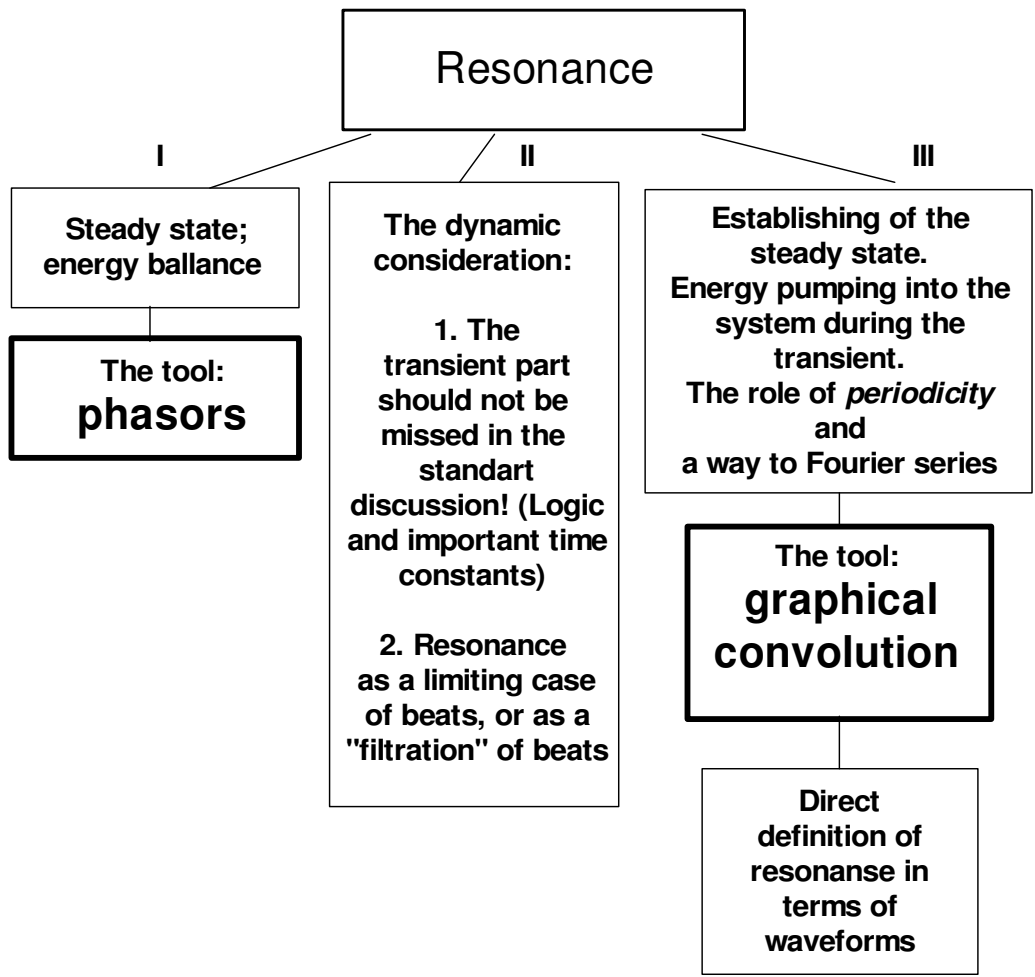

Fig. 5: The methodological points regarding the study of resonance in the present work.

Our notations need minor comments. As is customary in electrical engineering, the notation for $\sqrt{-1}$ is $j$. The small italic Latin 'v', $v$, is *voltage* in the *time domain* (i.e. a real value); $\hat{V}$ means *phasor*, i.e. a complex number in the frequency domain. $\lambda$ is the *dummy variable* of integration in a definite integral of the convolution type. It is measured in seconds, and the difference $t-\lambda$, where $t$ is time, often appears.

*1.2. The second-order equation*

The background formulae for both the *Q-t* and *Q-a* parts of the resonant response can be given by the Kirchhoff *voltage equation* for the electrical current $i(t)$ in a series *RLC* (resistor-inductor-capacitor) circuit driven from a source of sinusoidal voltage with amplitude $v_m$:

$$L\frac{di}{dt} + Ri + \frac{1}{C}\int i(t)dt = v_m \sin \omega t \ . \qquad (1)$$

Differentiating (1) and dividing by $L \neq 0$, we obtain

$$\frac{d^2 i}{dt^2} + 2\gamma \frac{di}{dt} + \omega_o^2 i(t) = \frac{\omega v_m}{L} \cos \omega t \qquad (2)$$

with the *damping factor* $\gamma = \dfrac{R}{2L}$ and the *resonant frequency* $\omega_o = \dfrac{1}{\sqrt{LC}}$.

For purely resonant excitation, the input sinusoidal function is at frequency $\omega = \omega_o$, or at a very close frequency, as defined in Section 2.

## 2. The steady state response (*Q-a* part)

Let us briefly recall the standard phasor (impedance) treatment of the final *Q-a* (steady state) part of a system's response. We can focus here only on the results associated with the amplitude, the phase relations follow straightforwardly from the expression for impedance [3,4].

In order to characterize the *Q-a* part of the response, we use the common notations of [3,4]: the damping factor of the response $\gamma \equiv \dfrac{R}{2L}$, the resonant frequency $\omega_o = \dfrac{1}{\sqrt{LC}}$, the quality factor

$$Q = \frac{\omega_o}{2\gamma} = \frac{\omega_o L}{R} = \frac{\sqrt{L/C}}{R}$$

and the frequency at which the system self-oscillates:

$$\omega_d \equiv \sqrt{\omega_o^2 - \gamma^2} \approx \omega_o - \frac{\gamma^2}{2\omega_o} = \omega_o(1 - \frac{1}{4Q^2}) \approx \omega_o \ .$$

Note that it is assumed that $4Q^2 \gg 1$ and thus $\omega_d$ and $\omega_o$ are practically indistinguishable. Thus, although we *never ignore $\gamma$ per se*, the much smaller value $\gamma/Q \sim \gamma^2/\omega_o$ can be ignored. When speaking about 'precise resonant excitation', we shall mean setting $\omega$ with *this* degree of precision, but when writing $\omega \neq \omega_o$, we shall mean that $\omega - \omega_o = O(\gamma)$, and not $O(\gamma/Q)$. Larger than $O(\gamma)$ deviations of $\omega$ from $\omega_o$ are irrelevant to the topic of resonance.

The impedance of the series circuit is $Z(j\omega) = R + j\omega L + \dfrac{1}{j\omega C}$, and the phasor approach simply gives the *amplitude* of the steady state solution of (2) as:

$$i_m(\omega) = |\hat{I}(j\omega)| = \left|\frac{\hat{V}}{Z(j\omega)}\right|$$

$$= \frac{|\hat{V}|}{\left|R + j\omega L + \dfrac{1}{j\omega C}\right|} = \frac{v_m}{\sqrt{R^2 + \left(\omega L - \dfrac{1}{\omega C}\right)^2}} \quad . \qquad (3)$$

For $\omega - \omega_o \ll \omega_o$, when $\omega^2 - \omega_o^2 \approx 2\omega_o(\omega - \omega_o) \approx 2\omega(\omega - \omega_o)$,

$$i_m(\omega) \approx \frac{v_m}{2L\sqrt{\gamma^2 + (\omega - \omega_o)^2}} \quad . \qquad (4)$$

From (4), the frequencies at "half-power level", for which $i(\omega) = \dfrac{1}{\sqrt{2}}(i_m)_{\max}$, are defined by the equality $(\omega - \omega_o)^2 = \gamma^2$, from which we obtain $\omega_1 = \omega_o - \gamma$ and $\omega_2 = \omega_o + \gamma$, i.e. for the circuit's frequency "pass-band" $\Delta\omega \equiv \omega_2 - \omega_1$ we have, with the precision taken in the derivation of (4), that $\Delta\omega = 2\gamma$.

It is remarkable that *however small is $\gamma$, it is easy, while working with the steady state, to detect differences of order $\gamma$ between $\omega$ and $\omega_o$, using the resonant curve/response described by (4)*.

Figure 6 illustrates the resonance curve.

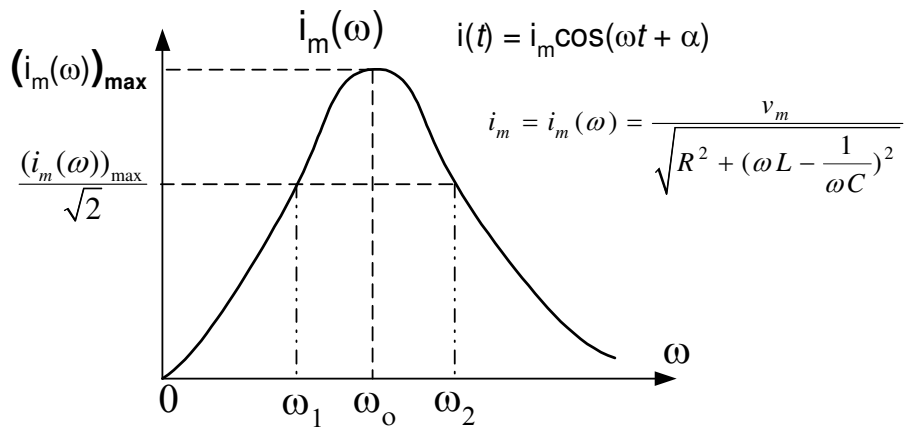

Fig. 6: The resonance curve.  $\Delta\omega \equiv \omega_2 - \omega_1 = 2\gamma$;  $Q = \dfrac{\omega_o}{\Delta\omega}$.

Though this figure is well known, it is usually not stressed that since each point of the curve corresponds to some steady-state, a certain time is needed for the system to pass on from one point of the curve to another one, and the sharper the resonance the more time is needed. The physical process is such that just for a small $\gamma$ the establishment of this response takes a (long) time of the order of

$$\frac{1}{\gamma} = \frac{\omega_o}{\gamma} \frac{1}{\omega_o} = 2Q \frac{T_o}{2\pi} = \frac{1}{\pi} QT_o \sim QT_o, \quad (T_o = \frac{2\pi}{\omega_o}) \quad (5)$$

which is not directly seen from the resonance curve.

The relation $1/\gamma \sim QT_o$ for the transient period should be remembered regarding *any* application of the resonance curve, in any technical device. The case of a mistake caused by assuming a quicker performance for measuring input frequency by means of passing on from one steady state to another, is mentioned in [2]. This mistake is associated with using only the resonance curve, i.e. thinking only in terms of the frequency response.

### 3. The standard time-domain outlook

The full solution of equation (2) can be explicitly composed of *two* terms; the first, denoted as $i_h$, originates from the *homogeneous* (h) equation, and the second, denoted as $i_{fs}$, represents the *finally obtained* (fs) periodic oscillations, i.e. is the *simplest* (but not the only possible!) partial solution of the *forced* equation:

$$i(t) = i_h(t) + i_{fs}(t). \quad (6)$$

It is important that the zero initial conditions cannot be fitted by the second term in (6), $i_{fs}(t)$, continued backward in time to $t = 0$. (Indeed, no sinusoidal function satisfies both the conditions $f(t) = 0$ and $df/dt = 0$, at any point.) Thus, it is obvious that a nonzero term $i_h(t)$ is needed in (6). This term is

$$i_h(t) = e^{-\gamma t}(K_1 \cos \omega_d t + K_2 \sin \omega_d t), \quad (7)$$

and at least one of the constants $K_1$ and $K_2$ is nonzero here.

Furthermore, it is obvious from (7) that the time needed for $i_h(t)$ to decay is of the order of $1/\gamma \sim QT_o$, which is (5). How*ever, according to the two-term structure of (6), the time needed for $i_{fs}(t)$ to be established, i.e. for $i(t)$ to become $i_{fs}(t)$ is just the time needed for $i_h(t)$ to decay. Thus, the established 'frequency response' is attained only after the significant time of order $QT_o \sim Q$.*

Unfortunately, this elementary logic argument following from (6) is missed in [3-5] and many other technical textbooks that ignore the *Q-t* part of the resonance and directly deal only with the *Q-a* part.

However form (6) is also not optimal here because it is not directly seen from it that for zero initial conditions not only $i_{fs}(t)$, but also the decaying $i_h(t)$ are directly proportional to the amplitude (or scaling factor) $v_m$ of the input wave.

*That is, the general form (6) by itself does not make obvious the fact that **when choosing zero initial conditions**, we make the response function **as a whole** (including the transient) to be a tool for studying the input function.*

It would be better to have *one* expression/term from which this feature of the response is well seen. This better formula appears below.

### 4. The use of graphical convolution

We pass on to the constructive point, -- the convolution integral presenting the resonant response, and its graphical treatment. It is desirable for a good "system understanding" of the topic that the concepts of *zero input response* (ZIR) and *zero state response* (ZSR), especially the latter one, be known to the reader.

Briefly, ZSR is the *partial response* of the circuit, which satisfies the zero initial conditions. As $t \to \infty$ (and only then), it becomes the final steady-steady response, i.e. becomes the *simplest* partial response (whose waveform can be often guessed).

Appendix illustrates the concepts of ZIR and ZSR in detail, using a first-order system and stressing the distinction between the forms ZIR + ZSR, and (6), of the response.

Our system-theory tools are now the *impulse* (*or shock*) *response h*(*t*) (*or Green's function*) and the integral response to $f_{inp}(t)$ for zero initial conditions.

$$f_{out}(t) = (h * f_{inp})(t) = \int_0^t h(\lambda) f_{inp}(t-\lambda) d\lambda = (f_{inp} * h)(t). \quad (8)$$

The convolution integral (8) is an example of ZSR, and it is the most suitable tool for understanding the resonant excitation.

It is clear that (contrary to (6)) the total response (8) is directly proportional to the amplitude of the input function.

Figure 7 shows our schematic system:

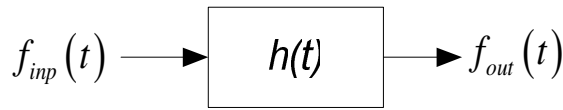

**Fig. 7**:    The input-output map ($f_{inp} \to f_{out}(t) = ZSR(t)$) given by 'impulse response' *h*(*t*).

In terms of Fig. 19 (see Appendix), $f_{inp}$ means a usual generator input.

Of course, the *system-theory outlook* does not relate only to electrical systems; this "block-diagram" can mean influence of a mechanical force on the position of a mass, or a pressure on a piston, or temperature at a point inside a gas, etc..

Note that if the initial conditions are zero, they are simply not mentioned. If the input-output map is defined solely by $h(t)$ (e.g. when one writes in the domain of Laplace variable $F_{out}(s) = H(s)F_{inp}(s)$), it is always ZSR.

In order to treat the convolution integral, it is useful to briefly recall the simple example [5] of the first order circuit influenced by a single square pulse. The involved physical functions are shown in Fig. 8, and the associated "*integrand situation*" of (8) is shown in Fig. 9.

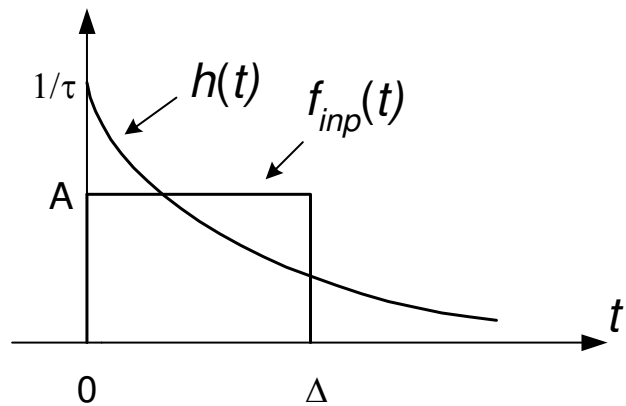

Fig. 8: The functions for the simplest example of convolution. (A first-order circuit with an input block-pulse.)

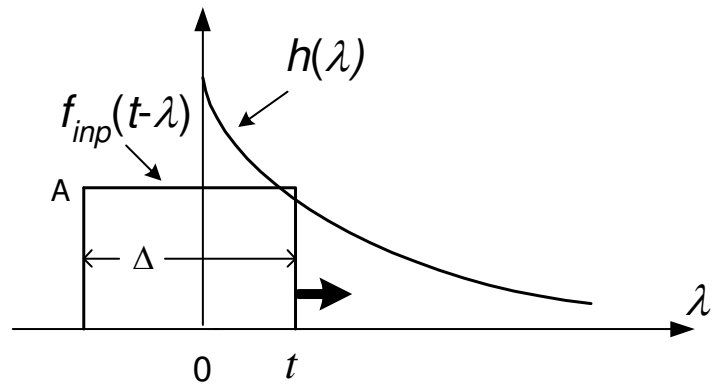

Fig. 9: The functions appearing in the integrand of the convolution integral (8). The "block" $f_{inp}(t-\lambda)$ is riding (being moved) to the right on the $\lambda$-axes, as time passes. We multiply the present curves in the interval $0 < \lambda < t$ and, according to (8), take the area under the result, in this interval. When $t < \Delta$, only the interval $(0,t)$ is relevant to (8). When $t > \Delta$, only the interval $(t-\Delta, t)$ is actually relevant, and because of the decay of $h(t)$, $f_{out}(t)$ becomes decaying, obviously.

It is *graphically obvious* from Fig. 9 that the maximal value of $f_{out}(t)$ is obtained for $t = \Delta$, when the rectangular pulse already fully overlaps with $h(\lambda)$, but still "catches" the initial (highest) part of $h(\lambda)$. This simple observation shows the strength of the graphical convolution for a qualitative analysis.

### *4.1 The* (*resonant*) *case of a sinusoidal input function acting on the second-order system*

For the second-order system with weak losses, we use for (8)

$$h(t) = \frac{\omega_o^2}{\omega_d} e^{-\gamma t} \sin \omega_d t \qquad . \qquad (9)$$
$$\sim e^{-\gamma t} \sin \omega_d t \approx e^{-\gamma t} \sin \omega_o t, \quad \gamma \ll \omega_o \ (Q \gg 1).$$

As before, we apply

$$f_{inp}(t) = f_m \sin \omega_d t \approx f_m \sin \omega_o t \,.$$

Figure 10 builds the solution (8) step by step; first our $h(\lambda)$ and $f_{inp}(t-\lambda)$ (compare to Fig. 9), then the product of these functions, and finally the integral, i.e. $f_{out}(t) = S(t)$.

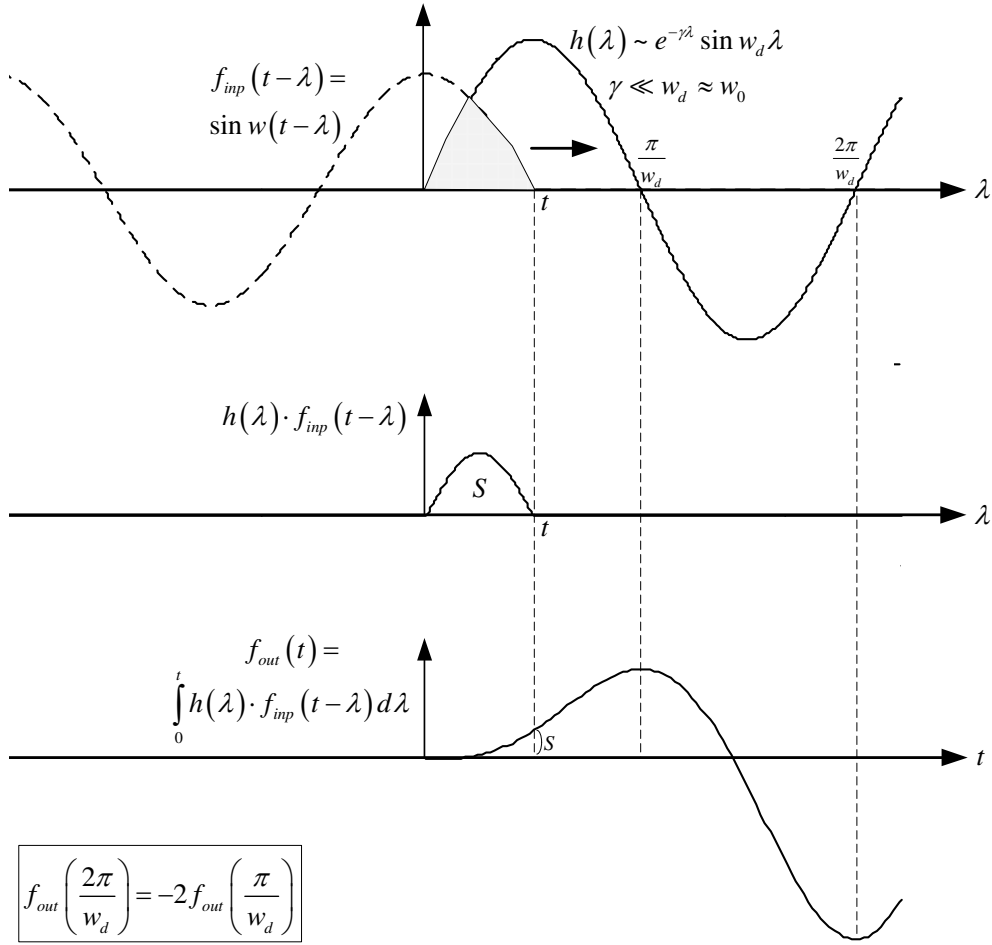

Fig. 10: Graphically obtaining of the resonant response for a second-order oscillatory system and a sinusoidal input, according to (8). The envelope (not shown) has to pass via the maxima and minima of $f_{out}(t)$ appearing in the last graph.

On the upper graph, the 'train' $f_{inp}(t-\lambda)$ travels to the right, starting at $t = 0$, on the middle graph we have the integrand of (8). The area $f_{inp}(t) = S(t)$ under the integrand's curve appears as the final result on the third graph.

The extreme values of $S(t)$ are $S(k\frac{\pi}{\omega_d})$, obviously. For $k$ odd these are positive maxima because the overlaps in the upper drawing are then '+' with '+', and '–' with '–'. For $k$ even these are negative minima because we multiply the opposite polarities in the overlap $f_{inp}(t-\lambda)h(\lambda)$ each time. Thus, $S(\frac{\pi}{\omega_d}) > 0$ and $S(2\frac{\pi}{\omega_d}) < 0$.

In view of the basic role of the overlapping of $f_{inp}(t-\lambda)$ with $h(\lambda)$, it is worthwhile to look forward a little and compare Fig. 10 to Figs. 14-15 that relate to the case of an input *square wave*.  For the upper border of integration in (8) be $t = k\dfrac{\pi}{\omega_d}$ and for very weak damping of $h(\lambda)$ the situations being compared are very similar.  The distinction is that in order to obtain the extremes of $f_{inp}(t)$, we integrate in Fig. 15 the absolute value of several *sinusoidal pieces* (*half-waves*), while in Fig. 10 we integrate the *squared sinusoidal pieces.*  As the point, since we integrate, in each case, $k$ similar pieces (all positive, giving a maximum of $f_{out}(t)$ and all negative, giving a minimum), the results of such an integration appear to be directly proportional to $k$.

Thus, if $\gamma = 0$, when $h(\lambda)$ is periodic, from the *periodic nature* of also $f_{inp}(t)$ it follows that

$$f_{out}(k\frac{\pi}{\omega_d}) \sim (-1)^{k+1} k \sim k , \qquad (10)$$

for any integer $k$, which is a linear increase in envelope for the two very different input waves, in the spirit of Fig. 1.

For a small but finite $\gamma$, $0 < \gamma \ll \omega_o$, the initial linear increase has high precision only for some first few $k$ when $t \sim T_o \sim 1/\omega_o \ll 1/\gamma$, i.e. $\gamma t \ll 1$, or $e^{-\gamma t} \approx 1$. (The damping of $h(t)$ may be ignored *for these k*.)

Observe that the finally obtained periodicity of $f_{out}(t)$ follows only from that of $f_{inp}(t)$, while the linear increase requires periodicity of both $f_{inp}(t)$ and $h(t)$.

The above discussion suggests the following simplification of the impulse response of the circuit, useful for analysis of the resonant systems.  This simplification is a useful preparation for the rest of the analysis.

### *4.2 A simplified h(t) and the associated envelope of the oscillations*

Considering that the parameter $1/\gamma$ appears in the above (and in Fig.3) as some symbolic border for the linearity, let us take a constructive step by suggesting a geometrically clearer situation when this border is artificially made sharp by introducing an idealization/simplification of *h(t)*, which will be denoted as $h_S(t)$.

In this idealization, -- that seems to be no less reasonable and suitable in qualitative analysis than the usual use of the vague expression "*somewhere at 't' of order* $1/\gamma$", -- we replace *h(t)* by a finite "piece" of non-damping oscillations of total length $1/\gamma$.

We thus consider that however weak the damping of *h(t)* is, for sufficiently large *t*, when $t \gg 1/\gamma \sim QT_o$, we have $e^{-\gamma t} \ll 1$, i.e. the oscillations become strongly damped with respect to the first oscillation. For $t > 1/\gamma$ the further 'movement' of the function $f_{inp}(t-\lambda)$ to the right (see Fig. 10 again) becomes less effective; the exponentially decreasing tail of the oscillating *h(t)* influences (8), via the overlap, more and more weakly, and as $t \to \infty$, $f_{out}(t)$ stops to increase and becomes periodic, obviously.

We simplify this qualitative vision of the process by assuming that up to $t = 1/\gamma$, there is no damping of *h(t)*, but, starting from $t = 1/\gamma$, *h(t)* completely disappears.

That is, we replace the function $e^{-\gamma t}\sin\omega_d t$ by the function $h_S(t) = [u(t) - u(t - 1/\gamma)]\sin\omega_o t$ where $u(t)$ is the unit step function. The factor $u(t) - u(t - 1/\gamma)$ here is a "cutting window" for $\sin\omega_o t$. This is the formal writing of the "piece" of the non-damping self oscillations of the oscillator. See Fig. 11.

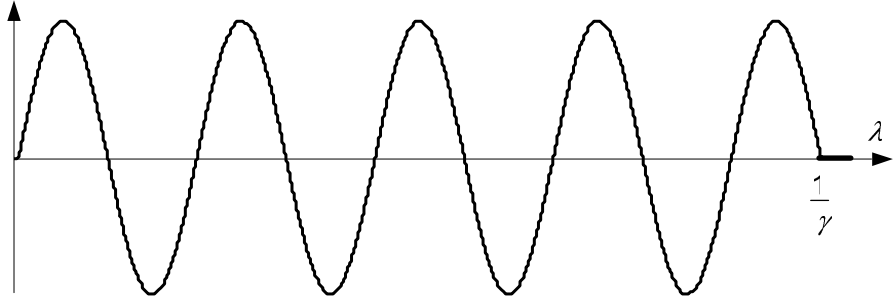

<u>Fig. 11</u>: The simplified $h(t)$ (named $h_S(t)$): there is no damping at $0 < t < 1/\gamma$, but for $t > 1/\gamma$ it is identically zero, i.e. we first ignore the damping of the real $h(t)$, and then cut it completely. This idealization expresses the undoubted fact that the interval $0 < t < 1/\gamma$ is dominant, and makes the treatment simpler. A small change in $1/\gamma$ which makes the oscillatory part more pleasing by including in it just the (closest) integer number of the half waves, as shown here, may be allowed, and when using $h_S(t)$ below we shall assume for simplicity that the situation is such.

For $h_S(t)$, it is obvious that when the "train" $f_{inp}(t-\lambda)$ crosses in Fig. 10 the point $t = 1/\gamma$, the graphical construction of (8), i.e. $f_{out}(t)$, becomes a periodic procedure. Figuratively speaking, we can compare $h_S(t)$ with a railway station near which the infinite train $f_{inp}(t-\lambda)$ passes; some wagons go away, but similar new enter, and the total overlapping is repeated periodically.

The same is also analytically obvious, since when setting, for $t > 1/\gamma$, the upper limit of integration in (8) as $1/\gamma$, we have, because of the periodicity of $f_{inp}(\cdot)$, the integral:

$$f_{out}(t > \frac{1}{\gamma}) = \int_0^{1/\gamma} h(\lambda) f_{inp}(t-\lambda)\, d\lambda \qquad (11)$$

as a periodic function of $t$.

As is illustrated by Fig. 12, -- which is an approximation to the envelope shown in Fig.3, -- the envelope of the output oscillations become completely saturated for $t > 1/\gamma$.

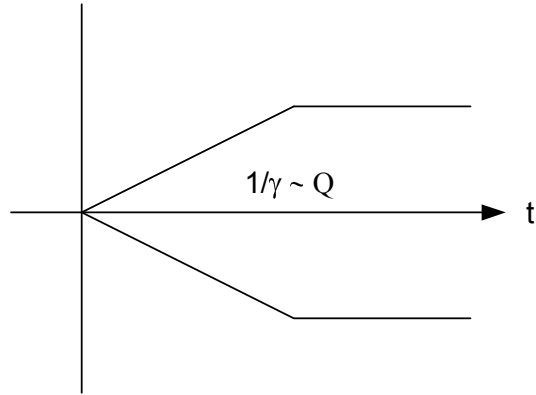

Fig. 12: The envelope of $f_{out}(t)$ obtained for the simplified $h(t)$ shown in Fig. 11.

Fig.12 clearly shows that both the amplitude of the finally established steady-state oscillations and the time needed for establishing these oscillations are proportional to $Q$, while the initial slope is obviously independent of $Q$.

It is important that $h_S(t)$ can be also constructed for more complicated functions $h(t)$ (for which it may be, for instance $h(t+T/2) \neq -h(t)$) and also then the graphical convolution is easier formulated in terms of $h_S(t)$. As an example relevant to the theoretical investigations, -- presenting the maximal values of the established oscillations, obtained for $t_k \gg 1/\gamma$,

$$|f_{out}(t_k)| = \left| \int_0^\infty f_{inp}(t_k - \lambda) h(\lambda) d\lambda \right|, \quad t_k \gg 1/\gamma,$$

as

$$|f_{out}(\gamma^{-1})| = \left| \int_0^{1/\gamma} f_{inp}(\gamma^{-1} - \lambda) h_S(\lambda) d\lambda \right|,$$

we can easily reduce, using periodicity of $f_{inp}(t)$ for any oscillatory $h(t)$ (and $h_S(t)$), the contribution of the interval $(0, 1/\gamma)$, to that of a small interval, as was $(0, \frac{\pi}{\omega_d})$ in the above.

### 4.3 Non-sinusoidal input waves

The advantage of the graphical convolution is not so much in the calculation aspect. It is an easy for imagination (insight) procedure, and it is a flexible tool in the qualitative analysis of the time processes. The graphical procedure makes it absolutely clear that the really basic point for a resonant response is not sinusoidality, but *periodicity* of the input function. Not being derived from the spectral (Fourier) approach, this observation heuristically completes this approach, and may be used (see below) in an introduction to Fourier analysis.

Thus, let us now take $f_{inp}(t)$ as the rectangular wave shown in Fig. 13,

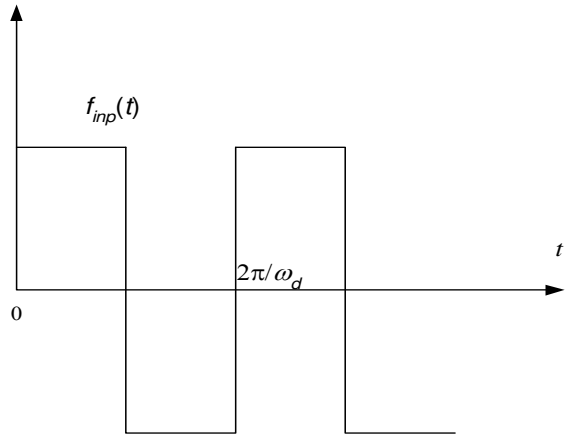

Fig. 13: The rectangular wave at the input.

and follow the way of Figs. 9 and 10, in the sequential Figs 14 and15.

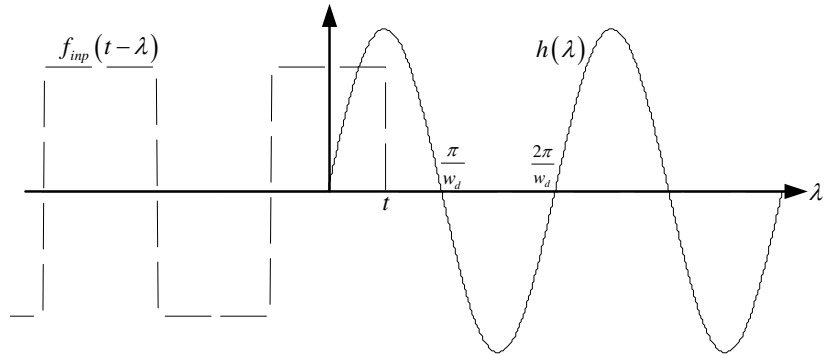

Fig. 14: Convolution with a rectangular wave at the input.  Compare to Figs. 9 and 10.

Here too, the envelope of the resonant oscillations can be well outlined by considering $f_{out}(t)$ at instances $t_k = \dfrac{k\pi}{\omega_d}$; first of all $\dfrac{\pi}{\omega_d}$, $\dfrac{2\pi}{\omega_d}$, and $\dfrac{3\pi}{\omega_d}$, for which we respectively have the first maximum, the first minimum, and the second maximum of $f_{out}(t)$.

There are absolutely the same qualitative (geometric) reasons for resonance here, and Fig. 15 explains that if the damping of $h(t)$ is weak, i.e. some first sequential half-waves of $f_{inp}(t-\lambda)h(t)$ are similar, then the respective extreme values of $S(t) = f_{out}(t)$ form a linear increase in envelope.

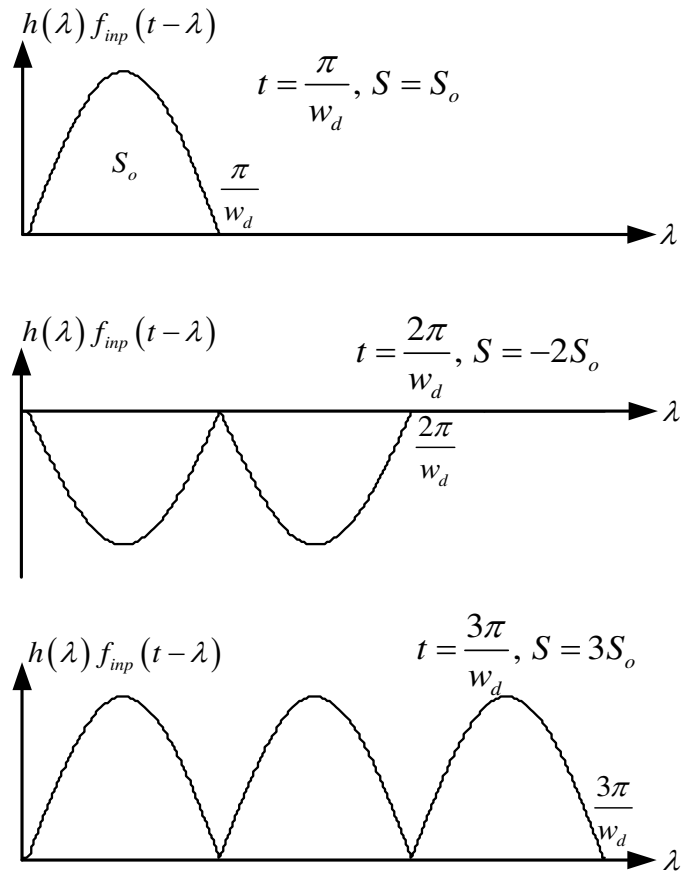

Fig. 15: Continuation of the creation of the convolution value, after Fig. 14. The function $h(\lambda) f_{inp}(t-\lambda)$ is shown at three intervals $0 < \lambda < t = k\dfrac{\pi}{\omega_d} \approx k\dfrac{\pi}{\omega_o}$ , $k = 1,2,3$, for which the area under this function of $t$ has local extremes. $S_o = S(2\pi/\omega_d))$ denotes the area under a half-wave of $h(\lambda) f_{inp}(t-\lambda)$. Damping of $h(t)$ is ignored and we have here the cases of $S = S_o$, $S = -2S_o$, and $S = 3S_o$, which represent the output function at its extremes, see Fig. 16.

Figure 16 shows $f_{out}(t) = S(t)$ at these extreme points.

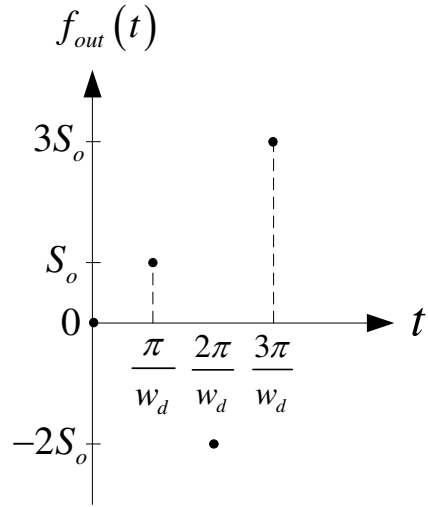

Fig. 16: Linear increase of the envelope (ideal in the in the lossless situation) for the square wave input. Compare to Figs. 1, 3 and 12.

Though it is not easy to find the precise $f_{out}(t)$ everywhere, for the envelope of the oscillations, which passes through the extreme points, the resonant increase in the response amplitude is absolutely clear.

Figures 10 and 14-16 make it clear that many other waveforms with the correct period would likewise cause resonance in the circuit. Furthermore, for the overlapping to remain good, we can change not only $f_{inp}(t)$, but also $h(t)$. Making the form of the impulse response more complicated means making the system's structure more complicated, and thus graphical convolution is also a valuable starting point for studying resonance in complicated systems in terms of the waveforms. This point of view will be realized in Section 5 where we generalize the concept of resonance.

Thus, using the algorithm of the graphical convolution, we make two more methodological steps; a pedagogical one in Section 4.4, and the constructive one in Section 5.

### *4.4  Let us try to "discover" the Fourier series*

The conclusion regarding the possibility of obtaining resonance using a non-sinusoidal input optimistically means that when swinging a swing with a child, the father need not develop a sinusoidal force. Moreover, the nonsinusoidal input even has some obvious advantages. While the sinusoidal input wave leads to resonance only when its frequency has the correct value, exciting resonance by means of a non-sinusoidal wave can be done at different basic frequencies (kick the swing not at every oscillation), which is, of course, associated with Fourier series.

Let us see how, using graphical convolution, we can reveal harmonic structure of a function, *still not knowing anything about Fourier series*. For that, let us continue with the case of square wave input, but take now such a waveform with a period that is 3 times longer than the period of self oscillations of the oscillator. Consider Fig. 17.

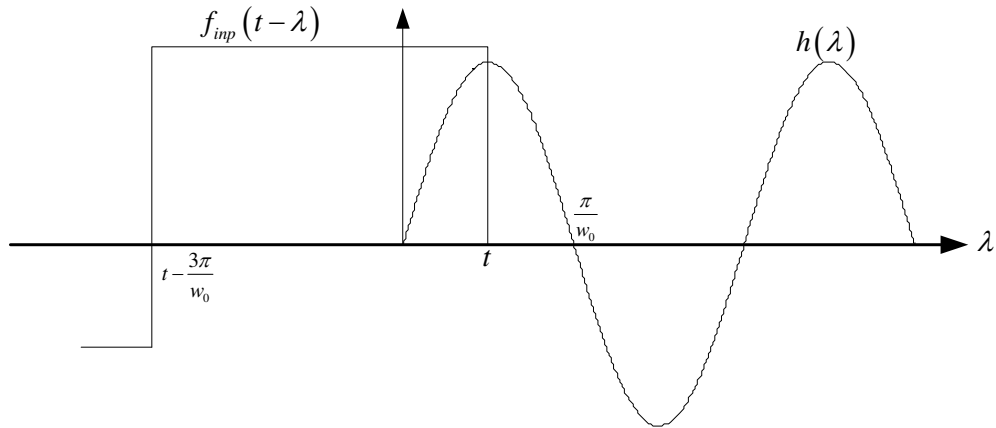

Fig. 17: We "discover" the Fourier series using graphical convolution. The convolution of $h(t)$ with the square wave having $T = 3T_o$.

This time, the more distant instances, $t = \dfrac{3\pi}{\omega_d}, \dfrac{6\pi}{\omega_d}$, and $\dfrac{9\pi}{\omega_d}$, are obviously most suitable for understanding how the envelope of the oscillations looks like.

One sees that also for $T = 3T_o$, *the same* geometric 'resonant mechanism' exists, but the transfer from $T = T_o$ to $T = 3T_o$ makes the excitation significantly less intensive. Indeed, see Fig. 18 comparing the present extreme case of $t = \dfrac{3\pi}{\omega_d}$ to the extreme case of $t = \dfrac{\pi}{\omega_d}$ of Fig. 15.

We see that each extreme-overlap is now only *one third* as effective as was the respective maximum overlap in the previous case. That is, at $t = \dfrac{3\pi}{\omega_d}$ we now have what we previously had at $t = \dfrac{\pi}{\omega_d}$, which means a much slower increase in the amplitude in time.

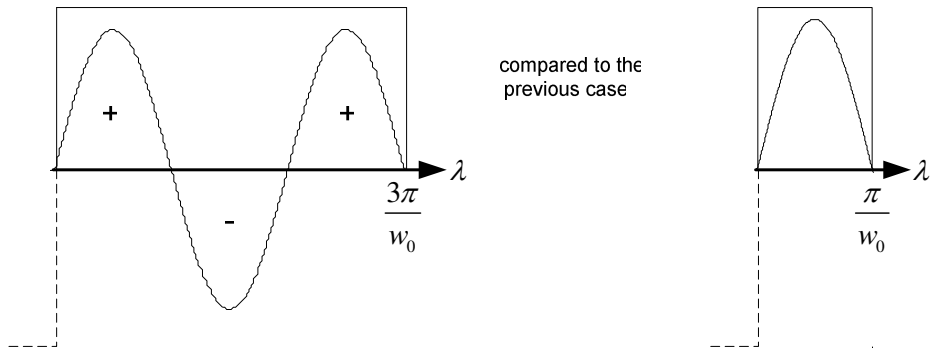

Fig. 18: Because of the mutual compensation of two half-waves of $h(\lambda)$, only each third half-wave of $h(t)$ contributes to the extreme value of $f_{out}(t)$, and the maximum overlaps between $h(\lambda)$ and $f_{inp}(t)$ are now one third as effective as before. The Reader is asked (this will soon be needed!) to similarly consider the cases of $T = 5T_o$, etc..

   Since $f_{out}(t)$ is now increased at a much slower rate, but $1/\gamma$ is the same (i.e. the transient lasts the same time), the amplitude of the final periodic oscillations is respectively smaller, which means weaker resonance in terms of frequency response.
   Let us compare the two cases of the square wave studied to the initial case of the sinusoidal function. The case of the "non-stretched" square wave corresponds to the input $\sin \omega_o t$, while according to the conclusions derived Fig. 18, the case of the "stretched" wave corresponds to the input $\frac{1}{3}\sin \omega_o t$. We thus simply (and roughly) reduce the change in period of the non-sinusoidal function to the equivalent change in amplitude of the sinusoidal function.
   Let us now try, -- as a tribute to Joseph Fourier, -- to speak not about the same circuit influenced by different waves, but about the same wave influencing different circuits. Instead of increasing $T$, we could decrease $T_o$, thus testing the ability of the same square wave to cause resonance in the different oscillatory circuits. For the new circuit, the graphical procedure remains the same, obviously, and the ratio 1/3 of the resonant amplitudes in the compared cases of $T/T_o = 3$ and $T/T_o = 1$ remains. However, since while decreasing $T_o$ from $T$ to $T/3$, we increase the resonant frequency of the circuit from $\omega_o$ to $3\omega_o$, the respective associations become not with the inputs $\sin \omega_o t$ and $\frac{1}{3}\sin \omega_o t$, but with the inputs $\sin \omega_o t$ and $\frac{1}{3}\sin 3\omega_o t$.
   In fact, we are thus testing the square wave using *two* simple oscillatory circuits of different self-frequencies. Namely, connecting in parallel to the source of the square wave voltage two simple oscillatory circuits with self frequencies $\omega_o$ and $3\omega_o$, we reveal for one of them the action of the square wave as that of $\sin \omega_o t$, and for the other as that of $\frac{1}{3}\sin 3\omega_o t$.
   This associates the square wave of height *A*, with the series

$$f(t) \sim A(\sin \omega t + \frac{1}{3}\sin 3\omega t + \frac{1}{5}\sin 5\omega t ...) \qquad (12)$$

(which precisely is $f = (4A/\pi)(\sin \omega t + \ldots\ )$ ).

   Let us check this result by using the arguments in the inverse order. The first sinusoidal term of series (12) roughly corresponds to the square wave with $T = T_o$ (i.e. $\omega = \omega_o$), and in order to make the *second term* resonant, we have to change the self-frequency of the circuit to $\omega_o = 3\omega$, i.e. make $\omega = \frac{1}{3}\omega_o$, or $T = 3T_o$, which just is our second "experiment" in which the reduced to 1/3 intensity of the resonant oscillations is indeed obtained, in agreement with (12).

   It is possible to similarly graphically analyze a triangular wave at the input, or a sequence of periodic pulses of an arbitrary form (more suitable for the father kicking the swing) with a period that is an integer of $T_o$.

   One notes that such figures as Fig. 18 are relevant to the standard integral-form of Fourier coefficients. However on the way of graphical convolution, this similarity arises *only* for the extremes $(f_{out}(t))_{\max} = |f_{out}(t_k)|$, and this way is independent.

## 5. A generalization of the definition of resonance, in terms of mutual adjustment of $f_{inp}(t)$ and $h(t)$

   After working out the examples of the graphical convolution, we are now in position to formulate a wider *t*-domain definition of resonance.

   In terms of the graphical convolution, the analytical symmetry of (8):

$$(h * f_{inp})(t) = (f_{inp} * h)(t) \qquad (13)$$

means that besides observing the overlapping of $f_{inp}(t-\lambda)$ and $h(\lambda)$, we can observe overlapping of $h(t-\lambda)$ and $f_{inp}(\lambda)$. In the latter case, the graph of $h(-\lambda)$ starts to move to the right at $t = 0$, as was in the case with $f_{inp}(-\lambda)$.

   Though equality (13) is a very simple mathematical fact, similar to the equalities $ab = ba$ and $(\vec{a},\vec{b}) = (\vec{b},\vec{a})$, in the context of graphical convolution there is a non-triviality in the *motivation* given by (13), because the possibility to move $h(-\lambda)$ also suggests changing the *form* of $h(\cdot)$, i.e. starting to deal with a *complicated system (or structure)* to be resonantly excited. We thus shall try to define resonance, i.e. the optimization of the peaks of $f_{out}(t)$ (or its *r.m.s* value), in the terms of more arbitrary waveforms of $h(t)$, while the case of the sinusoidal $h(\cdot)$, i.e. of simple oscillator, appears as a particular one.

***5.1. The optimization of the overlapping of*** $f(\lambda) \equiv f_{inp}(t-\lambda)$ ***and*** $h(\lambda)$ ***in a finite interval, and creation of the optimal periodic*** $f_{inp}(t)$

Let us continue to assume that the losses in the system are small, i.e. that $h(t)$ is decaying so slowly that we can speak about at least few oscillatory spikes (labeled by $k$) through which the envelope of the oscillations passes during its linear increase.

Using notation $S_o$ of Figs 15 and 16, we speak about the extreme points of the graph of the resulting function $f_{out}(t)$, i.e., about the points whose coordinates are $(t_k, f_{out}(t_k))$, or

$$(t_k, (-1)^{k+1} S_o k), \quad k = 1, 2, \ldots . \quad (14)$$

In view of the examples studied, the extreme points of $f_{out}(t)$ are obtained when $t_k$ are the zero-crossings of $h(t)$, because only thus the overlapping of $f_{inp}(t_k - \lambda)$ with $h(\lambda)$ can be made maximal.

Comment: Assuming that the parameters of the type $\gamma/\omega_0$ of the different harmonic components of $h(t)$ are different, one sees that for a non-sinusoidal damping $h(t)$, the distribution of the zero-crossings of $h(t)$ can be changed with the decay of this function, and thus for a periodic $f_{inp}(t)$ the condition

$$sign[f_{inp}(t_k - t)] = sign[h(t)] ,$$

or

$$sign[f_{inp}(t_k - t)] = -sign[h(t)] ,$$

considered for $k \gg 1$ need not be satisfied in the whole interval of the integration ($0 < \lambda < t_k$) related to the case of $t_k \gg T$. However since both the amplitude-type decays, and the change in the intervals between the zeros are defined by the same very small damping parameters, the resulted effects of imprecision are of the same smallness. Both problems are not faced at all when we use the "generating interval" and employ $h_S(t)$ instead of the precise $h(t)$. The fact that any use of $h_S(t)$ is anyway associated with error of order $Q^{-1} \sim \gamma T_o$ points at the expected precision of the generalized definition of resonance.

Thus, $\{t_k\}$, measured with respect to the time origin, i.e. with respect to the moment when $f_{inp}(t)$ and $h(t)$ arise, is assumed be given by the known $h(t)$. Of course, we assume the system to be an oscillatory one, for the parameters $t_k$ and $S_o$ of our graphical constructions to be meaningful.

Having the linearly increasing sequence $|f_{out}(t_k)| = S_o k$ belonging to the envelope of the oscillations, and wishing to increase the finally established oscillations, we have to increase the factor $S_o$, obviously.

However since $S_o$ and the whole intensity of $f_{out}(t)$ can be increased not only by the proper wave-form of $f_{inp}(.)$, but also by an amplitude-type *scaling factor*, for the general discussion some *norm* for $f_{inp}(.)$, has to be introduced.

For the definitions of the norm and the scalar products of the functions appearing during adjustment of $f_{inp}(t)$ to $h(t)$, it is sufficient to consider a *certain* (for a fixed, not too large $k$) *interval* $(t_k, t_{k+1})$, -- the one in which we can calculate $S_o$. This interval can be simply $(0, t_1)$, or $(0, T)$.

The norm over the chosen interval is taken as

$$\| f \| = \sqrt{\int_{t_k}^{t_{k+1}} f^2(t) dt} \ . \tag{15}$$

For instance, $\| \sin \omega t \|$ calculated over interval $(0, \frac{T}{2} = \frac{\pi}{\omega})$, or $(\frac{\pi}{\omega}, \frac{2\pi}{\omega})$ is $\sqrt{\frac{\pi}{2\omega}}$, as is easy to find, by using the equality $\sin^2 \alpha = \frac{1}{2}(1 - \cos 2\alpha)$.

Respectively, the scalar product of two functions is taken as

$$(f_1, f_2) = \int_{t_k}^{t_{k+1}} f_1(t) f_2(t) dt \ . \tag{16}$$

With these definitions, the set of functions defined for the purpose of the optimization in the interval $(t_k, t_{k+1})$ forms an (infinite-dimensional) Euclidean space.

For the quantities that interest us, we have from (16) for the absolute values:

$$S_o = |(f, h)| = \left| \int_{t_k}^{t_{k+1}} f(t) h(t) dt \right|, \tag{17}$$

where (see Figs 10,14 and 15 ) it is set for simplicity of writing

$$f(t) \equiv f_{inp}(t_{k+1} - t), \quad t_k < t < t_{k+1} \ . \tag{18}$$

That we do not ascribe to "$f(\cdot)$" index '$k$' is justified by the fact that the *particular* interval $(t_k, t_{k+1})$ to be used is finally chosen very naturally.

The basic relation $|f_{out}(t_k)| = S_o k$ means that any local extremum of $f_{out}(t)$ is a sum of such scalar products as (16).

Observe that the physical dimensions of $\| \cdot \|$ and $(\cdot, \cdot)$ are

$$[\| \cdot \|] = [f][t^{1/2}] = \frac{[f]}{[\omega^{1/2}]}, \tag{19}$$

and

$$[(\cdot, \cdot)] = [f_1][f_2][t] = \frac{[f_1][f_2]}{[\omega]}. \tag{20}$$

Observe also from (15,16) that

$$(f,f) = \| f \|^2 , \qquad (21)$$

and that if we take $f_2 \sim f_1$, i.e.

$$f_2(t) = K f_1(t), \quad K \in \mathbb{R}, \qquad (22)$$

then (21) is generalized to

$$(f_1, f_2) = sign[K] \| f_1 \| \cdot \| f_2 \|.$$

Indeed, using (21), and then the obvious equalities $K = |K| sign[K]$ and $|K| \cdot \| f \| = \| Kf \|$, we obtain:

$$\begin{aligned}(f_1, f_2) &= (f_1, K f_1) = K(f_1, f_1) = K \| f_1 \|^2 \\ &= sign[K] |K| \cdot \| f_1 \| \cdot \| f_1 \| = sign[K] \| f_1 \| \cdot \| K f_1 \| \\ &= sign[K] \| f_1 \| \cdot \| f_2 \|\end{aligned} \qquad (23)$$

The factor $sign[K]$ means, in particular, that excitation of an oscillatory circuit can be equivalently done by either an $f_{inp}(t)$ or $-f_{inp}(t)$. (Consider the concept of "overlapping" in this view.)

It follows from (23) that if (22) is provided, then

$$|(f_1, f_2)| = \| f_1 \| \cdot \| f_2 \| \quad \text{(for (22) provided)} \qquad (24)$$

Furthermore, we use that the following general inequality

$$|(f_1, f_2)| \leq \| f_1 \| \cdot \| f_2 \| \qquad (25)$$

takes place. In view of (15,16), (25) is just the known Cauchy-Bunyakovsky integral inequality.

Comparing (25) with (24), we see that condition (22) provides optimization of $|(f_1, f_2)|$. Applied to $f$ and $h$, i.e. to $S_o = |(f,h)|$, this conclusion re optimization says that the condition $f \sim h$ optimizes $S_o$. Thus, $f \sim h$ optimizes the extremes $f_{out}(t_k) \sim S_o k$ of the system's response.

Thus, we finally have two points:

(a) We consider the proper interval $(t_k, t_{k+1})$ using which we create the periodic $f_{inp}(t)$.

(b) The proportionality $f \sim h$ in this interval is the optimal case of the influence an oscillatory circuit by $f_{inp}(t)$.

Items (a) and (b) are our definition of the generalized resonance. The case of sinusoidal $h(t)$ is obviously included since the proportionality to $h(t)$ requires $f(t)$ to also be sinusoidal, of the same period.

This mathematical situation is the constructive point, but it appeared as a too "dry" one, and the discussion below (Sections 5.3 and 5.5) of the optimization of $S_o$ from a more physical point of view is useful, leading us to very compact formulation of the extended resonance condition.

However, let us first of all use the simple oscillator checking how essential is the direct proportionality of $f$ to $h$, i.e. what may be the quantitative miss when the waveform of $f(t)$ differs from that of $h(t)$ in the chosen interval $(t_k, t_{k+1})$.

### *5.2. An example for a simple oscillator*

Let us compare the cases of the *square* (Figs 13-15) and *sinusoidal* (Fig. 10) input waves of the same period, for $S_o$ defined in the interval $(0, \frac{T}{2} = \frac{\pi}{\omega})$. Of course, the norms of the input functions have to be equal for the comparison of the respective responses. (Note that in the consideration of the above figures, equality of the norms was *not* provided, and thus the following result cannot be derived from the previous discussions.)

Let the height of the square wave be 1. Then, $f_{inp}^2(\pi - \lambda) = 1$ everywhere, and according to (15), the norm is obtained as $\sqrt{\frac{\pi}{\omega}}$. For obtaining the same norm for a sinusoidal input, we write it as $K \sin \omega t$ and find $K > 0$ so that

$$\| K \sin \omega t \| = \sqrt{\frac{\pi}{\omega}} \ ,$$

i.e.

$$K = \frac{\sqrt{\pi}}{\sqrt{\omega} \, \| \sin \omega t \|} \ .$$

Because of the symmetry of the sinusoidal and square-wave inputs, in both cases $f_{inp}(\lambda) = f_{inp}(t - \lambda) \equiv f(\lambda)$ in the interval $(0, \frac{\pi}{\omega})$. For either of the input waveforms the norm of $f_{inp}(t)$ now equals $\sqrt{\frac{\pi}{\omega}}$, and for $h(t) = \sin \omega t$ of the simple oscillator (the damping in this interval is ignored), we have, according to (17) and (25),

$$S_o = (f, h) \leq \left( \sqrt{\frac{\pi}{\omega}} \right) \| \sin \omega t \| = \sqrt{\frac{\pi}{\omega}} \sqrt{\frac{\pi}{2\omega}} = \frac{\pi}{\omega \sqrt{2}} \approx \frac{2.221}{\omega} \ ,$$

as the upper bund.

While for the response to the square wave we have

$$S_o = \int_0^{\pi/\omega} 1 \cdot \sin \omega t \, dt = \frac{1}{\omega} \int_0^{\pi} 1 \cdot \sin x \, dx = \frac{2}{\omega}$$

*only*, for the response to the input $K \sin \omega t$ we have, for the $K$ found ,

$$S_o = \int_0^{\pi/\omega} \left( \frac{\sqrt{\pi}}{\sqrt{\omega}\,\|\sin\omega t\|} \sin\omega t \right) \sin\omega t\, dt$$

(26)

$$= \frac{\sqrt{\pi}}{\sqrt{\omega}\,\|\sin\omega t\|} \|\sin\omega t\|^2 = \sqrt{\frac{\pi}{\omega}}\,\|\sin\omega t\|$$

which just is the maximal possible value $\|f\|\cdot\|h\|$ for $S_o$, as it *has to be* because in this case $f(\lambda) = f_{inp}(\frac{\pi}{\omega}-\lambda) \sim h(\lambda)$ in the interval $(0 < \lambda < \frac{\pi}{\omega})$.

The "relative *missing* the optimality" in the case of the square wave, which we wanted to found, is

$$\left|\frac{2-2.221}{2.221}100\%\right| = |-9.95\%| \approx 10\%.$$

As the next step of our discussion let us consider the condition $f \sim h$, by interpreting (22,24,25) in the usual physical Euclidean space when: (a) the *uniqueness* of the condition (24) is obvious, and (b) preserving the norms during obtaining the proportionality of the vectors, is not any problem.

### *5.3. Analogy with the usual vectors*

In the mathematical sense, the set of functions that can be used for the optimization of $S_o$ is analogous to the set of usual vectors.

For the scalar product $(\vec{a},\vec{b})$ of two usual vectors $\vec{a}$ and $\vec{b}$ we have (compare to (25))

$$|(\vec{a},\vec{b})| \leq |\vec{a}\|\vec{b}| \qquad (27)$$

(meaning "$\cos\theta \leq 1$") where the *equality* is obtained only when the vectors are mutually proportional ("$\theta" = 0$), i.e. similarly directed:

$$\vec{a} \sim \vec{b}\ . \qquad (28)$$

The latter relation is *obvious*, in particular because it is obvious that while rotation of the usual vector (say a pencil) when directing it in parallel to another vector (another pencil), the length of this vector is unchanged. This point is much more delicate regarding the norm of a function being adjusted to $h(t)$, which is the "rotation" of the "vector" in the function space. Since the waveform of the function is being changed, its norm can be also changed.

Thus, the usual physical space *very simply* gives the extreme value of $|(\vec{a},\vec{b})|$ as $|\vec{a}\|\vec{b}|$:

$$\max|(\vec{a},\vec{b})| = |\vec{a}\|\vec{b}|$$

(this to be compared to the longer derivation of (24)), and we shall know this maximal value numerically if $|\vec{a}|$ and $|\vec{b}|$ are given.

Since our "vectors" are the time functions, and the functional analog of (28) is

$$f(t) \sim h(t) ,$$

we very simply obtain, *by the mathematical equivalence of the function and the vector spaces*, condition (24), i.e. that only an $f(t)$ that is directly proportional to $h(t)$ can give an extreme value for $S_o$.

For the vectors of the same length (e.g. for unit vectors) $|\vec{b}|=|\vec{a}|$, and the condition of optimality, $\vec{a} \sim \vec{b}$, becomes $\vec{a} = \pm \vec{b}$. In the functional space, the latter means that if $\|f(t)\| = \|h(t)\|$, then in order to have $S_o$ maximal we should take $f(t) = \pm h(t)$.

### *5.4 Comments*

Though, in fact, the condition $f = Kh$ is a *generalization* of the condition $\omega = \omega_o$ relevant to purely sinusoidal functions, seeing the latter particular case as especially important, one can consider $f \sim h$ to be a direct *analogy* to the condition $\omega = \omega_o$ of the standard definitions of [1,3-5]. Then, both of the equalities, $\omega = \omega_o$ and $f = Kh$, appear in the associated theories as sufficient conditions for obtaining resonance in a linear oscillatory system. The norms become important at the next step, namely regarding the theoretical conditions of system's linearity, which always include some limitations on intensity of the function/process, and the application. For applications, the real properties of the physical source of $f_{inp}(t)$ (e.g. a voltage source) whose *power* will here be proportional to $K^2$, obviously require $K$ to be limited.

The requirement of preserving the norm $\|f(t)\|$ during realization of $f(t) \sim h(t)$ also necessarily originates from the practically useful formulation of the resonance problem as the *optimization problem* that requires *calculation* of the optimized peaks (or rms value) of $f_{inp}(t)$.

If $h(t+T/2) \neq -h(t)$, then the interval in which the scalar products (i.e. the Euclidean functional space) are defined, has to be taken over the whole period of $h(t)$, i.e. as $S_o = \int_0^T f_{inp}(T-\lambda)h(\lambda)d\lambda$. ($T = T_o$ is a necessary condition.) The general procedure thus is that after concluding for the first period of $h(t)$ that $f_{inp}(T-\lambda) \sim h(\lambda)$, i.e. having $f_{inp}(t) \sim h(T-t)$ for $0 < t < T$ (this is the same; replace $T-\lambda$ by a single symbol, in the integral), we continue $f_{inp}(t)$ periodically for $t > T$.

The interval in which we define $S_o$ can be named the "generating" interval.

We can finally write the optimal $f_{out}(t)$ resulted from the optimal $f_{inp}(t)$ as:

$$f_{out}(t) = \int_0^t h(\lambda) f_{inp}(t - \lambda) d\lambda$$

$$= \int_0^t h(\lambda) h_{(0,T) periodic}(\lambda) d\lambda \qquad (29)$$

where the function $h_{(0,T) periodic}(t)$ is $h(t)$ in the generating interval, periodically continued for $t > T$.

We turn now to an informal, but a strong "physical abstraction", suggested by the comparison of the two Euclidean spaces. This abstraction leads us to a very compact formulation of the generalized definition of resonance.

### 5.5. *A symmetry argument for formulation of the generalized definition of resonance*

For the usual vector space, we have well developed vectorial analysis, in which *symmetry arguments* are widely employed. The mathematical equivalence of the two spaces under consideration suggests that such arguments, -- as far as they related to the scalar products, -- are legitimized also in the functional space.

Recall the simple field problem in which the scalar field (e.g., electrical potential)

$$\varphi(\vec{r}) = (\vec{a}, \vec{r}) \qquad (30)$$

is given by means of a constant vector $\vec{a}$, and it is asked in what direction to go in order to have the steepest change of $\varphi(\vec{r})$.

As the methodological point, -- one need *not* know how to calculate gradient. It is just *obvious* that only $\vec{a}$, or a *proportional vector*, can show the direction of the gradient, since there is only *one fixed vector given*, and it is simply impossible to "construct" from the given data any other constant vector, defining another direction for the gradient.

We thus consider the *axial symmetry* introduced by $\vec{a}$ in the physical space that can be seen *ad hoc* as the "space of the radius-vectors", and conclude that while catching the steepest increases of $(\vec{a}, \vec{r})$, we must go with some $\vec{r} \sim \vec{a}$.

Let us compare this very lucid situation with that of the functional space. In the problem of making the envelope of the *convolution* $f_{out}(t)$ (for the whole interval $0 < \lambda < t$ ) be increasing as steep as possible, we have, in view of the relation $|f_{out}(t_k)| = S_o k$, to optimize the scalar product $S_o = (f, h)$. This is quite similar to (30), because here $h(t)$ is the only fixed "vector" involved, -- i.e. no other "directions" in the functional space are given.

Thus, by the direct analogy to the fact that the gradient must be proportional to $\vec{a}$, the optimal $f(t)$ must be proportional to $h(t)$.

We thus can say that ***in terms of ZSR, i.e. in terms of the convolution integral response, resonance is a use of (or "obeying") the axial symmetry introduced by $h(t)$ in the space of the functions convolving with $h(t)$***.

This argument makes the generalized definition compact, easy to remember. One just should not forget that we optimize the factor $S_o$ in a certain interval, say the first period of *h*(*t*).

## 6. Discussion

The traditional teaching of resonance in technical textbooks in terms of a purely steady state (frequency) response and phasors, not deepening into the time process, i.e. into the *establishment* of the 'frequency response', is seen to be unsatisfactory.

The general tendency of engineering teachers to work only in the frequency domain, is explained, but is not justified by the importance of the fields of communication and signal processing. A good understanding of the time processes is needed in physics, chemistry, biology and also power electronics. We hope that the use of convolution integral suggested here can, to some extent, close any such logical gap when it appears, and can make the topic of resonance more interesting to a student. The described graphical application of convolution is also important for understanding the convolution integral per se. Last but not least, we hope that our generalized definition of resonance in terms of optimization of a scalar product in an interval will be useful.

On the way to the generalized definition, our hero was the father swinging a swing, and not the definitions of [1] and [3,4]. Everything relevant (even the Fourier series) can be directly understood from the *freedom* that the father has when enhancing the swing's oscillations.

In the historical plane, the simplicity of the mathematical treatment of the sinusoidal case once defined the general point of view on resonance and the standard classroom treatment, but we see that the convolution integral has became a sufficiently simple and common tool to make this definition wider.

The present criticism of the usual teaching resonance well correlates with the 'old' pedagogical advice by Ernst Guillemin [10] not to hurry with the frequency-domain analysis and to let the physical reality first be well understood in the time domain.

Direct study of waveforms (not necessarily using the graphical convolution) also reveals some specific resonant effects that are *not obtained at all* for a sinusoidal input [7-8]. Thus, for some rectangular-wave periodic input waves, a *resonant suppression* of the response-oscillations of a simple oscillator can occur *at certain, periodically repeated time intervals*, and only a direct analysis of the *waveforms* reveals this suppression [7-9]. It appears that the singularity of the waveform and its symmetry [7-9], and not Fourier (spectral) representation, reveal these "pauses" in the oscillatory function. Remarkably, since singularity and symmetry aspects are applied also to a nonlinear oscillatory circuit, these "pauses" in the oscillations can be similarly simply explained [7-9] for such a nonlinear circuit.

The topic of resonance is an important scientific and pedagogical point from which different mathematical and physical interpretations can be developed, and it should be revisited by a Teacher. Hopefully, the present study can help one in that.

## **Appendix:** The representation of the circuit response as ZIR(*t*) + ZSR(*t*) (some basic system-theory terminology for physicists)

Besides the standard mathematical representation (6) of the solution of a linear equation, system theory commonly uses another representation in which the output function is composed of a *Zero Input Response* (ZIR) and a *Zero State Response* (ZSR).

The ZSR is influenced by the generator inputs and satisfies *zero initial conditions* (this is the meaning of the words 'zero state'), and the ZIR is defined *only* by nonzero initial conditions, i.e. is *not* influenced by the generator's' inputs, which is the meaning of the words "zero-input" response.

Since both the generator-type input functions and the initial conditions can be defined freely, they are both legitimized inputs and altogether form a *generalized input*.

### *A1. The superposition with respect to the generalized input*

The concept of *generalized input* (Fig. 19), fully explain the construction of ZIR and ZSR via the superposition. Indeed, in the classical way of (6), $i_h(t)$ is found from the homogeneous equation which is *not* the given one; but is artificially introduced. That is, the determining of $i_h(t)$ is an *auxiliary problem* in which the generators inputs (that define the right-hand side of the given equation) are zero. The concept of generalized input requires doing *the same* also for the initial conditions, i.e. to additionally use the given equation with the artificially introduced initial zero conditions. Thus, according to the two different groups of the inputs we have two parts of the whole solution, obtained from the following auxiliary *independently solvable* problems:

For ZIR:   Homogeneous equation (zero generator inputs) plus the needed initial conditions.

For ZSR:   Given equation with zero initial conditions.

Figure 19 schematically illustrates this presentation of the linear response.

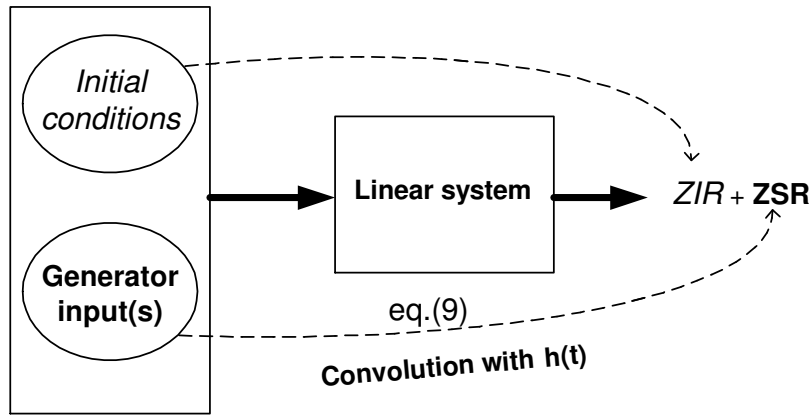

Fig. 19: The *generalized input* includes both generator inputs and initial conditions. (This becomes trivial when Laplace transform is used to transform a system *scheme*.) Considering the superposition of the output function of a linear system *for such an input*, we obtain the associated structure of the solution of the linear system in a form that is different from (6). Respectively, the output function is written as ZIR($t$) + ZSR($t$).

Figure 7 reduces Fig.19 to what we actually need for the processes with zero initial conditions. The logical advantage of the presentation ZIR + ZSR over (6) becomes clear in the terms of the superposition.

The ZSR includes both the needed for satisfying zero initial conditions damping transient and the final steady state given in its general form by the integral (A4) below. The oscillations shown in Fig. 3 are examples of ZSR.

The separation of the solution function into ZIR and ZSR is advantageous, e.g., when the circuit is used to analyze the input signal, i.e. when we wish to work only with the ZSR, when nonzero initial conditions can only bother.

The convolution integral (8) is ZSR, and, when speaking about system with constant parameters having one input and one output, the Laplace transform of ZSR($t$) equals $H(s)F_{inp}(s)$ where $H(s)$ is the 'transfer function' of the system, i.e. the Laplace transform of $h(t)$. *Each time when we speak about transfer function, we speak about ZSR i.e. zero initial conditions*.

It is easy to write $F_{out}(s)$ for our problem. Using the known formula for Laplace transform of periodic function, and setting the optimal $f_{inp}(T-\lambda) \equiv -h(\lambda)$, $\lambda \in [0,T]$, i.e. $f_{inp}(t) \equiv -h(T-t)$, $t \in [0,T]$, where $T$ is the period (in the sense of the generating interval) of $h(t)$, we have, for the periodically continued $f_{inp}(t)$, the Laplace transform of our $f_{out}(t)$ as (see (29))

$$F_{out}(s) = H(s)F_{inp}(s) = H(s)\frac{\int_0^T f_{inp}(t)e^{-st}dt}{1-e^{-sT}}$$

$$= H(s)\frac{-\int_0^T h(T-t)e^{-st}dt}{1-e^{-sT}} = H(s)\frac{\int_0^T h(t)e^{st}dt}{1-e^{sT}}$$

(the integration only over *the first* period and, finally, '+s' everywhere), which is relevant to different oscillatory $h(t)$.

### A.2 Example

Consider for $t > 0$ the following simplest example of the first-order system/equation

$$\frac{dy}{dt} + ay(t) = 0, \quad y(0) \neq 0 \text{ is given,}$$

where *a* and *A* are constants. Here, the solution of type (6), $i_h(t) + i_{fs}(t)$, is first $K\exp(-at) + A/a$, and when involving the initial condition, finally,

$$y(t) = (y(0) - A/a)e^{-at} + A/a, \quad t > 0, \qquad (A1)$$

with the initial conditions and the generator function "mixed" in the first term.

The ZIR + ZSR representation is obtained by rewriting this expression as:

$$y(t) = y(0)e^{-at} + \frac{A}{a}(1 - e^{-at}), \quad t > 0. \qquad (A1a)$$

The first term depends on the initial condition, i.e. is ZIR, and the second term depends on the generator input $Au(t)$ ($u(t)$ is the unit-step function) i.e. is ZSR.

It is easy to check that, as was said, ZIR can be *independently* found from the equation $y' + ay = 0$, and the given initial condition, and ZSR can be *independently* found from the given equation $y' + ay = A$, and the zero initial condition.

For $y(0) = 0$, $y(t) = ZSR(t) = \frac{A}{a}(1 - e^{-at})$, which can be also written as

$$y(t) = \int_0^t e^{-a(t-\lambda)}A\,d\lambda, \qquad (A2)$$

i.e. (as (8)) as

$$y(t) = \int_0^t h(t-\lambda)f_{inp}(\lambda)\,d\lambda, \qquad (A3)$$

where $h(t) = e^{-at}u(t)$ is the impulse response of the first-order circuit.

Considering (A1a), one sees that the ZSR includes (as $t \to \infty$) decaying components of the same *type* as the ZIR, and that the asymptotic response $A/a$ originates from the ZSR as $t \to \infty$, and *not at all* from the ZIR.

### A.3. $f_{out}(t)$ as $t \to \infty$

If (as in the above example) $f_{out}(\infty)$ exists, then it is obtained as

$$f_{out}(\infty) = \lim_{t \to \infty} \int_0^t h(t-\lambda) f_{in}(\lambda) d\lambda,$$

but if $f_{out}(\infty)$ does not exist, then as $t \to \infty$ the time function of the final state is given by making the upper limit of the integration infinity:

$$f_{out}(t) \underset{t \to \infty}{\sim} \int_0^\infty h(t-\lambda) f_{in}(\lambda) d\lambda \qquad (A4)$$

(that is, the roles of the argument '$t$' in (A3) are different for the different places in which it appears).

The integral in (A4) can be rewritten as

$$\int_{-\infty}^t h(\lambda) f_{in}(t-\lambda) d\lambda . \qquad (A5)$$

Dealing with the asymptotic solution (A5) is typical for *stochastic problems*, where, contrary to our statement of the resonance problem, the initial conditions are not important.

When speaking about convolution only in the form

$$\int_{-\infty}^\infty h(\lambda) f_{in}(t-\lambda) d\lambda$$

one misses the effects of the initial conditions which are important for system analysis, and it is inevitable that only the spectral approach appears be relevant.

### A.4. A case when ZIR + ZSR is directly obtained

When a differential equation can be directly solved by integration, the solution is directly obtained in the form of ZIR + ZSR. Thus, for Newton's equation written in the usual notations,

$$m \frac{d\vec{v}}{dt} = \vec{F}(t), \qquad (A6)$$

we have

$$\vec{v}(t) = \vec{v}(0) + \frac{1}{m}\int_0^t \vec{F}(\lambda)\,d\lambda \qquad (A7)$$

which obviously is ZIR + ZSR.  Superposition *with respect to the force* $\vec{F}(t)$ is realized only by the ZSR.

The easiness in finding the ZIR-part is explained by the fact that ZSR is found very straightforwardly.  Consider also

$$m\frac{d^2\vec{r}}{dt^2} = \vec{F}(t)$$

for $\vec{r}(0)$ and $\frac{d\vec{r}}{dt}(0)$ given.

The presentation ZIR + ZSR is generally relevant to *linear time-variant* (LTV, "parametric") equations that include the equations with constant parameters as a special case.  For instance, if the mass in (A6) depends on time, the integrand in ZSR in (A7) would be $\vec{F}(\lambda)/m(\lambda)$.  Generally, LTV equations are very difficult, but for *any linear homogenous equation (e.g. equation of parametric resonance)*, for which ZSR need not be found, *it follows from the linearity* that the solution (which then is just *ZIR*) has the form

$$\begin{aligned}y(t) = ZIR(t) &= F(y(0), y'(0),...,t)\\ &= y(0)f_1(t) + y'(0)f_2(t) + ...\end{aligned}$$

with all the functions known.  Since $y(0), y'(0),...$ are legitimized inputs, this is the usual linear superposition.